\documentstyle[preprint,aps]{revtex}

\def\e{\epsilon}

\newcommand{\be}{\begin{equation}}
\newcommand{\ee}{\end{equation}}

\input epsf

\begin{document}
\preprint{WISC-MILW-95-TH-17}
\title{Hawking Radiation and Unitary Evolution}
\author{Sukanta Bose,\footnote{Electronic address: {\em bose@csd.uwm.edu}}
Leonard Parker,\footnote{Electronic address: {\em leonard@cosmos.phys.uwm.edu}}
and Yoav Peleg\footnote{Electronic address: {\em yoav@csd.uwm.edu}} }
\address{Department of Physics \\
University of Wisconsin-Milwaukee, P.O.Box 413 \\
Milwaukee, Wisconsin 53201, USA}
\maketitle
\vskip 0.5cm
\begin{abstract}
We find a family of exact solutions to the semi-classical equations
(including back-reaction) of two-dimensional dilaton gravity,
describing infalling null matter that
becomes outgoing and returns to infinity without forming a black hole.
When a
black hole almost forms, the radiation reaching infinity in advance
of the original outgoing
null matter has the properties of Hawking radiation.
The radiation reaching infinity after the null matter consists
of a brief burst of negative energy that preserves unitarity
and transfers information faster than the theoretical bound for
positive energy.
\end{abstract}

\newpage

Recently we presented a modified two-dimensional (2D) dilaton gravity theory
that is exactly solvable semiclassically \cite{Bose}.
In this 2D theory (as in 4D Einstein gravity), infalling
null matter forms a black hole only if its energy $M$ is
above a certain critical value $M_{cr}$.
In Ref. \cite{Bose} we studied the
supercritical case, $M > M_{cr}$, in
which a black hole is formed and evaporates
by emitting Hawking radiation. In this work we study the subcritical
case, $M < M_{cr}$, in which
the infalling matter becomes outgoing and escapes to infinity
without forming a black hole.
Because we can obtain the numerical solution to the future of the
classical outgoing matter, we can see how correlations among outgoing created
particles preserve unitarity and how Hawking radiation originates as the
energy of the infalling matter approaches $M_{cr}$.

The effective action of the modified 2D dilaton gravity theory is \cite{Bose}
  \begin{eqnarray} \label{action}
S &=& {1\over 2\pi} \int d^2x \sqrt{-g(x)} \left[ ( e^{-2\phi} - \kappa \phi)
R(x) \right. \\
&+& (4 e^{-2\phi} + \kappa) (\nabla \phi)^2 + 4 \lambda^2 e^{-2\phi}
- {1\over 2} \sum_{i=1}^{N} \left. (\nabla f_{i})^2 \right]
\nonumber \\
&-&  {\kappa \over 8 \pi} \int \! \! d^2x \sqrt{-g(x)} \!
\int \! \! d^2x' \sqrt{-g(x')} R(x) G(x,x') R(x')  \nonumber ,
  \end{eqnarray}
where $\phi$ is the dilaton field, $R(x)$ is the 2D Ricci scalar,
$\lambda$ is a constant, $f_{i}$ are $N$ matter (massless scalar) fields,
$\kappa = \hbar N / 12$,
and $G(x,x')$ is an appropriate Green function for $\nabla^2$
(for more details see Ref. \cite{Bose}).
In null coordinates, $z^{\pm}$, and a conformal gauge,
$g_{++}=g_{--}=0$, $ g_{+-}={1\over 2} e^{2\rho}$,
the equations of motion (from varying $f_i$, $\phi$  and $g_{+-}$)
take the form
$\partial_{+}\partial_{-} Y = 0 = \partial_{+} \partial_{-} f_{i}$ and
$\partial_{+}\partial_{-} X = -\lambda^2 e^{-2Y}$,
where $X \equiv e^{-2\phi}$ and $Y \equiv \phi - \rho$.
The constraints (from varying $g_{++}$ and $g_{--}$) are
$-\partial_{\pm}^2 X - 2 \partial_{\pm}X \partial_{\pm} Y
-(T^f_{\pm\pm})_{c\ell} + \kappa \left[ (\partial_{\pm} Y)^2 +
\partial_{\pm}^2 Y + t_{\pm}(z^{\pm}) \right] = 0 $,
where $(T^f_{\pm\pm})_{c\ell} = {1\over 2} \sum_{i=1}^N (\partial_{\pm}
f_{i})^2$ is the classical (zero order in $\hbar$) contribution to the
energy-momentum tensor of the $f_{i}$ matter fields,
and $t_{\pm}(z^{\pm})$ are integration functions determined by the
specific quantum state of the matter scalar fields \cite{Bose}.
Because $\phi$ is a scalar, but $\rho$ is a function of $g_{+-}$ and
$\partial_+ \partial_- Y = 0$, it is possible to define coordinates
$x^+(z^+)$ and $x^-(z^-)$ in which $Y \equiv \phi - \rho$ is $0$.
In this work we use these ``Kruskal" coordinates $x^+, x^-$.

A general solution in the Kruskal gauge for the $X$-field equation
of motion and constraints is
  \begin{eqnarray} \label{generalsolution}
X(x^+,x^-) &=& -\lambda^2 x^+ x^-
- \int^{x^+} \! \! dx^+_2 \! \int^{x^+_2} \! \! dx^+_1
\left[ (T^f_{++})_{c\ell} - \kappa t_{+}(x^+_1) \right] \nonumber \\
& & -  \int^{x^-} \! \! dx^-_2 \! \int^{x^-_2} \! \! dx^-_1
\left[ (T^f_{--})_{c\ell} - \kappa t_{-}(x^-_1) \right] .
  \end{eqnarray}
The vacuum solutions are the ones for which $(T^f_{\pm\pm})_{c\ell}=
0$ and $t_{\pm}(x^{\pm})= (2x^{\pm})^{-2}$ \cite{Bose}.
They are static solutions because in the asymptotically
flat coordinates $\sigma^{\pm} = t \pm \sigma = \pm \lambda^{-1}
\mbox{log}(\pm \lambda x^{\pm})$, they take the $t$-independent form
$X_{vac}(\sigma) =
e^{2\lambda \sigma} - \kappa \lambda \sigma / 2 + C $,
where $C$ is a constant.
For $C > C^{*} \equiv (\kappa / 4)(\mbox{log}(\kappa /4) - 1)$
the vacuum solutions have a null singularity at $x^{\pm}=0$,
which is a finite affine distance from any point in the interior of the
space-time.
For $C<C^*$ the vacuum solutions have a time-like (naked) singularity
on the curve $\sigma = \sigma_{s}$, for which $X_{vac}(\sigma_{s})
=0$. The solution with $C=C^*$ is a semi-infinite throat which is everywhere
regular and geodesically complete \cite{Bose}.

In addition to initial conditions giving $(T^f_{++})_{c\ell}$ and $t_{+}(x^+)$
on asymptotic past null infinity, $\Im^-$, we must also determine
$(T^f_{--})_{c\ell}$ and $t_{-}(x^-)$. To do so, we choose
reflecting boundary conditions on the matter fields $f_i$
\cite{Fulling,Strominger1,Cartliz,Verlinde}.
We consider very localized infalling matter for which
$(T^f_{++})_{c\ell} (x^+)$ is non-zero only for $| x^+ - x^+_0| < \e$,
with $\lambda \e << 1$ (see  Fig. \ref{fig1}).
We take $t_+(x^+) = (2 x^+)^{-2}$ everywhere, which corresponds
to no quantum radiation on $\Im^-$.

The reflecting boundary conditions are imposed at the boundary curve on which
$X(x^+,x^-) \equiv X_{B}=$ constant. We take $X_{B} \geq 0$
so that the dilaton field $\phi = -{1\over 2} \mbox{log} (X)$ is real.
For a dynamical boundary curve, following a general trajectory
$x^+=x^+_{B}(x^-) \equiv p(x^-)$, the reflecting boundary condition is
\cite{Fulling,Strominger1,Cartliz,Verlinde}
  \be \label{dynamicalboundary}
T^f_{--} =
(p')^2  T^f_{++} + \kappa (p')^{1/2} \partial^2_{-} (p')^{-1/2} ,
  \ee
where $ ' = \partial / \partial x^-$,
and $T^f_{\mu \nu}$ is the energy-momentum tensor of the matter
fields, including both the classical and one-loop contributions. Namely,
$T^f_{\mu \nu} = (T^f_{\mu \nu})_{c\ell} + \langle T_{\mu \nu} \rangle$,
where $\langle T_{\pm \pm} \rangle = \kappa [ \partial^2_{\pm} \rho -
(\partial_{\pm} \rho)^2 - t_{\pm}(x^{\pm}) ] $
is the one-loop contribution to the energy-momentum tensor of the
scalar matter fields \cite{Davies}. The last term on the
right-hand-side of Eq. (\ref{dynamicalboundary}) arises because of
quantum particle creation from the boundary (which is effectively a
``moving mirror" \cite{Fulling}). The classical part
of $T^f_{\mu \nu}$ obeys $(T^f_{--})_{c\ell} = (p')^2 (T^f_{++})_{c\ell}$,
which is equivalent to the fields $f_{i}$ satisfying
either Neumann or Dirichlet boundary condition.
While the quantum part obeys $\langle T_{--} \rangle = (p')^2 \langle T_{++}
\rangle + \kappa (p')^{1/2} \partial^2_- (p')^{-1/2}$.
The boundary condition (\ref{dynamicalboundary}) is also conformally
invariant \cite{Cardy,Strominger1}.

To the past of the infalling null matter, $x^+ < x^+_0 - \e$, i.e.,
region I in Fig.\ \ref{fig1}, the geometry is one of the static vacuum
solutions.
The reflecting boundary condition
(\ref{dynamicalboundary}) implies that $(T^f_{--})_{c\ell}(x^-) = 0$
and $t_{-}(x^-) = (2 x^-)^{-2}$ for $x^- < x^-_B(x^+_0 - \e)$.
The solution (\ref{generalsolution}) to the past of the outgoing
classical null matter, regions I and II in Fig.\ \ref{fig1}, is therefore
  \be \label{bhsolution}
X_{<}(x^+,x^-) = - x^+ (\lambda^2 x^- + P_{+}(x^+))
- {\kappa\over 4} \mbox{log}(-\lambda^2 x^+ x^-) + {M(x^+)\over
\lambda} + C ,
  \ee
where $M(x^+) =
\lambda \int^{x^+} x^+_1 (T^f_{++})_{_{c\ell}}(x^+_1) dx^+_1$
and
$P_+(x^+) = \int^{x^+} (T^f_{++})_{_{c\ell}}(x^+_1) dx^+_1$ are the
mass and momentum of the classical infalling matter.


For initial static geometries
with $C < C^*$, the (very localized) infalling matter
forms a black hole only if $M > M_{cr} \equiv -\lambda^3 x^+_{0} x^-_{0} -
\kappa \lambda / 4$ , where $M=M(x^+ > x^+_0 + \e)$ is the
total mass of the infalling matter, and $x^-_{0}=x^-_{B}(x^+_{0})$.
The black hole evaporates by emitting Hawking radiation \cite{Bose},
and the semiclassical evolution seems to be non-unitary \cite{Bose,RST}.
In the subcritical case, $M < M_{cr}$, the classical infalling matter is
reflected from the boundary, which is always time-like (see Fig.\ \ref{fig1})
and the evolution is unitary.
We next find the solution numerically to the future of the outgoing classical
null matter (i.e., in region III of Fig.\ref{fig1}) in this subcritical case.

In Kruskal coordinates we use the constraint equations
to write the boundary condition (\ref{dynamicalboundary})
(on the one loop term) in the form
    \be \label{tboundary}
t_{-}(x^-) = (p')^2 t_{+}(p(x^-)) - { (p')^{1/2} \partial^2_- (p')^{-1/2}
\over 1 + \kappa/(2X_B)} \; .
  \ee
Since we are considering localized infalling
matter, the term
$(p')^{1/2} \partial^2_- (p')^{-1/2}$, which is of the order of
$M/(\lambda \e^2)$ in the region of classical reflection,
is large. However, in this work we consider the case in which
$X_B$ is small enough \cite{Yoav}
that $X_B M / (\kappa \lambda^3 \e^2) << 1$.
This is consistent with the condition, $\hbar/X_B << 1$,
that is necessary for the semiclassical
approximation to be valid everywhere,
since in the large $N$ limit we can take $\hbar$ to zero
sufficiently fast while keeping $\kappa = N\hbar/12$
constant \cite{Bose,Verlinde,Strominger1}.
In the limit $X_{B}M/( \kappa \lambda^3 \e^2) \rightarrow 0$,
the final term in Eq. (\ref{tboundary}) is negligible, and
Eq. (\ref{tboundary}) reduces to
$ t_{-}(x^-) = (p')^2 t_+(p(x^-)) = \left( p'/p \right)^2 \! /4 $.

One can study the solutions for general infalling matter
(not necessarily a very localized one).
The main features regarding correlations and Hawking radiation
with general infalling matter \cite{BPP2} are already present in
this case with $\lambda \e << 1$. In this limit
the solution to the future of $x^-_{0}$ (region III in Fig.\ \ref{fig1})
can be written in the form
  \be \label{futuresolution}
X_{>}(x^+,x^-) = - x^+(\lambda^2 x^- + P_{+})
- {\kappa\over 4} \mbox{log}(\lambda x^+) + F(x^-)  ,
  \ee
where $P_+=P_+(x^+>x^+_0+\e)$ is the total momentum of the classical
infalling matter and
the function $F(x^-)$ is determined by the boundary conditions.
{}From (\ref{futuresolution}) and the constraints, we
see that $\kappa t_{-}(x^-) = \partial^2_-F(x^-)$.
We define the dimensionless coordinate $y \equiv \lambda x^- + P_+/ \lambda $
and the dimensionless function $q(y) \equiv \lambda x^+_B(x^-(y))$, and
using (\ref{futuresolution})
get an ordinary (second order) differential equation for the boundary curve
$\lambda x^+_B = q(y)$
  \be \label{ordinaryequation}
\left[ y \; q(y) + {\kappa\over 4} \right] {d^2 q \over dy^2}  =
q(y) \left[ {\kappa\over 2} {1 \over q^2(y)} \left( {dq\over dy} \right)^2
- 2{dq\over dy} \right] .
  \ee
Define $y_0= \lambda x^+_0 + P_+ / \lambda $ to be
the value of $y$ just after the classical reflection (in the
limit $\lambda \e \rightarrow 0$).
Using the continuity of the solution and the classical reflecting
boundary condition, we get the initial data for $q(y)$:
$q(y_{0}) = \lambda x^+_0$ and $ (dq/dy)(y_{0}) =
\lambda x^+_0 \left[ M/(\lambda y_0) - \lambda x^+_0 - (4\lambda x^-_0)^{-1}
\right]/(\lambda x^+_0 y_0 + \kappa/4)$.
We choose $\lambda x^+_0 = - \lambda x^-_0 = 1$ and $\kappa = 10^{-6}$,
and solve (\ref{ordinaryequation}) numerically for different values
of the infalling mass.
The results that we get are qualitatively the same
for any value of $\kappa$ as long as $\kappa \lambda << M_{cr}$.
We use an embedded fifth order Runge-Kutta ODE integration routine
\cite{Press}.
The solutions for $q(y)$ are shown in Fig.\ \ref{fig2}.

%

When the infalling mass, $M$, is much less than $M_{cr}$,
as in Fig.\ \ref{fig2}a, the boundary curve is hardly affected by the
infalling matter, but when $M$
approaches $M_{cr}$, as in Fig.\ \ref{fig2}b, the timelike boundary
curve is strongly affected and approaches a null curve.

To see that the solutions are indeed stable (i.e., that the total
amount of energy radiated to $\Im^+$ is finite) we calculate the
created energy radiated to $\Im^+$,
$\langle T_{--} \rangle (u) = (\kappa \lambda^2 / 4) [ 1 -
4 y^2(u) t_{-}(y(u))]$, where $u=-\lambda^{-1} \mbox{log}(-y)$ and
$v = \lambda^{-1} \mbox{log}(\lambda x^+)$ are null coordinate in which
the metric is manifestly asymptotically flat (i.e., $\rho \rightarrow 0$
on $\Im^+$).
The results are shown in Fig.\ \ref{fig3} for the same
cases considered in Fig.\ \ref{fig2}.

%

The negative-energy radiation to the future of the original outgoing
null matter (i.e., in the region $u>u_{0}$) approaches
zero exponentially fast. The magnitude of the total radiated
negative energy is bounded by
$|\int_{u_0}^{\infty} \langle T_{--}(u) \rangle du | < (\kappa \lambda
/ 4) \mbox{log}(4M_{cr}/\lambda \kappa)$.
If $\Delta t$ is the time as measured by an asymptotic observer
in which this energy $E$ is radiated to $\Im^+$, then we obtain a quantum
inequality, $|E| \Delta t \sim \kappa  (M_{cr}-M)^2
\left( \mbox{log}[ M_{cr} /( M_{cr} - M)] \right)^2/(M M_{cr}) <
\kappa=N \hbar / 12$, of the type discussed by Ford and Roman \cite{Ford}.
The Heisenberg time-energy uncertainty principle implies that an attempt
to measure the energy of this burst in the available time $\Delta t$ will
disturb the energy by at least $|E|$.

We have verified numerically that the magnitude of negative energy
radiated to the future of the classical reflected matter ($u>u_0$) is equal
to that of the positive quantum radiation reaching $\Im^+$
before the classical reflected matter ($u<u_0$).
Thus energy is conserved. The total amount of
energy on $\Im^+$ is just that of the original reflected
null matter, $E^{tot}_{\Im^+} = \int_{-\infty}^{\infty} (T^f_{--})(u) du =
\int_{-\infty}^{\infty} (T^f_{--})_{c\ell} du = M$,
and the solutions are indeed stable.

In Fig.\ \ref{fig3}a the mass is far below the critical mass and the
quantum flux of
radiation is much less than that of thermal Hawking radiation, while
in Fig.\ \ref{fig3}b the flux of radiation before the classical
reflection (for $u$ in the range $u_0-\lambda^{-1} < u < u_0$)
approaches that of Hawking radiation.
As $M$ approaches the critical mass, the radiation before
the classical reflection becomes indistinguishable from Hawking radiation
originating from a black hole.

Since the semiclassical evolution of this reflecting solution
is unitary by construction
\cite{tHooft,Mikovic,Verlinde,Strominger1}, the radiation to the
future of $u_{0}$ must be strongly
correlated with the radiation to the past of $u_{0}$.
To see this explicitly we calculate the correlation function
  \be \label{correlationfunction}
C_{\mu \nu, \mu' \nu'}(x,x') \equiv \langle T_{\mu \nu}(x) T_{\mu' \nu'}(x')
\rangle - \langle T_{\mu \nu}(x) \rangle \langle T_{\mu' \nu'}(x') \rangle .
  \ee
We are especially interested in $C_{--,--}(u,u')$ which describes the
correlations in the outgoing radiation on $\Im^+$.
For reflecting boundary conditions there is a closed form expression
\cite{Cartliz} :
  \be \label{cuu}
C_{--,--}(u,u') = {1\over 8\pi^2} { \left( \partial_u v_{_{B}}(u) \right)^2
\left(\partial_{u'} v_{_{B}}(u') \right)^2
\over  \left( v_{_{B}}(u) - v_{_{B}}(u') \right)^4 } ,
  \ee
where $v_{_{B}}(u)$ is the boundary curve in the coordinates
 $v = \lambda^{-1} \mbox{log}(\lambda x^+)$ and $u = - \lambda^{-1} \mbox{log}
(-(\lambda x^- + P_+/\lambda))$.
Let us first consider the correlations in the radiation reaching $\Im^+$
before the classical reflected matter. For $u < u_{0}$ we have $v_{_{B}}(u) =
\lambda^{-1} \mbox{log}(-\lambda^2 x^+_0 x^-_0)
 - \lambda^{-1} \mbox{log} (e^{-\lambda u}  + P_{+}/\lambda)$,
and for $M$ near $M_{cr}$ we have $e^{\lambda u_{0} } >> 1$. Taking $u<u_0$ and
$u'<u_0$, with both near $u_{0}$, we find to
leading order in $\lambda e^{-\lambda u}/P_{+}$  and in
$\lambda e^{-\lambda u'}/P_+$,
  \be \label{thermalcuu}
C_{--,--}(u,u') = {\lambda^4\over 8\pi^2} {(e^{-\lambda u})^2
(e^{-\lambda u'})^2 \over (e^{-\lambda u} - e^{-\lambda u'})^4 }
\left( 1 + \mbox{O} ({\lambda e^{-\lambda u} \over P_{+}} ) \right) .
  \ee
The leading term in (\ref{thermalcuu}) is just the
correlation function in 2D for thermal black body
radiation \cite{Cartliz} with temperature $T = \lambda / 2\pi$,
the temperature of 2D dilatonic black holes \cite{Witten,CGHS}.
Through the same calculation that gave Eq. (\ref{thermalcuu}),
we find that (\ref{thermalcuu}) is also the correlation function for
Hawking radiation from an evaporating 2D black hole.

To explicitly verify
the correlations between the quantum radiation reaching $\Im^+$
before and after the classical reflected matter, we numerically calculate
(\ref{cuu}) by holding $u'$ fixed on
either side of $u_0$ and plotting $C_{--,--}(u,u')$ as a function of $u$.
The results are shown in Fig.\ \ref{fig4},
where the mass of the infalling matter is just below the critical mass,
$M = 0.99 M_{cr}$.

In Fig.\ \ref{fig4}a we show the correlations between the radiation at
$\lambda u' = 3.6$, and the radiation elsewhere on $\Im^+$.
This value of $\lambda u'$ is sufficiently close to $\lambda u_0 (=4.6)$
that the radiation has the same form as the Hawking radiation from
a black hole. For $u<u_0$ the correlation is almost
thermal, as in Eq. (\ref{thermalcuu}).
It diverges at $u=u'$ and becomes small for
$|u - u'| \sim 1/\lambda$. The most dramatic effect occurs
just after the reflected matter reaches $\Im^+$, i.e., at $u=u_{0} + \delta$.
The correlations rise (continuously when $\e$ is finite \cite{BPP2})
to extremely high values of order $e^{15}$. These
huge correlations are required for unitary evolution
because the negative-energy radiation reaching $\Im^+$
after the classical reflected matter is very localized in time.

%
%

Fig.\ \ref{fig4}b shows the correlations calculated numerically
between the negative-energy
radiation reaching $\Im^+$ just after
the classical reflected matter and the radiation elsewhere on $\Im^+$.
In this figure we take $\lambda u'= \lambda u_0 + \delta$ and $\delta
\rightarrow 0^+$.
The correlations diverge at $u=u'$ and decrease rapidly
when $u > u'$. On the other hand the correlations with the
Hawking radiation ($u<u_0$) remain extremely high for relatively
large values of $\lambda |u - u'|$. In contrast, if one were to take
$\lambda u' >> \lambda u_0$ then the only significant correlations
would occur for $\lambda u$ near $\lambda u'$.

The above results may plausibly be interpreted as arising from
the creation of particle-antiparticle pairs \cite{Parker}.
The particles reach infinity and give rise to the positive (Hawking)
radiation, while the antiparticles carrying negative energy
are reflected from the boundary and
give rise to the negative energy radiated to $\Im^+$
after the classical reflected matter.
If it were not for the negative-energy burst the correlations between
the particles and antiparticles would be lost and the final state would
be a mixed state. Thus, the negative-energy burst can be regarded as
carrying information equal to the magnitude of the entropy ${\cal S}$
of the Hawking radiation reaching $\Im^+$ ahead of the classical reflected
matter. As $M$ approaches $M_{cr}$, where $M_{cr} >> \kappa \lambda / 4 $,
we find that
${\cal S} \approx (N/ 12) \mbox{log} [4M_{cr}/(\kappa \lambda)]$.
The negative-energy burst thus carries information to $\Im^+$ at the
rate $\dot{I} = {\cal S} / \Delta t$. Using the earlier upper bound on
$|E| \Delta t$, we find that this gives $\dot{I} \geq (N \lambda
/ 48 ) ( 1 - M/M_{cr})^{-2} $. The theoretical upper
bound on the bulk rate of linear information flow in $N$ channels
for $E > 0$ is
\cite{Bekenstein} $\dot{I}_{max} \leq [E / (2\pi \hbar)] \mbox{log}_{2}(N)$
to within a factor of order 1. If this bound can be extended to $E < 0$
by replacing $E$ by $|E|$, then it gives for the present system:
$\dot{I}_{max}  \leq [N \lambda
/ (96 \pi)] \mbox{log} [4M_{cr}/(\kappa \lambda)] \mbox{log}_{2}(N)$.
This bound is clearly exceeded for large but finite $N$
in our system as $M$ approaches $M_{cr}$.

\vskip 5pt

We thank Bruce Allen, John Friedman, Jorma Louko and Eli Lubkin for helpful
discussions and especially Scott Koranda for help with the numerical analysis.
This work was supported by the National Science Foundation under grant
PHY 95-07740.

\newpage

\begin{figure}
\caption{Penrose diagram for a typical subcritical solution. The 2D space-time
is the interior of the 1D null hypersurfaces $\Im^-$, $\Im^+$ and the time-like
boundary curve, which intersect at $i^-=(x^+=0,x^-=-\infty )$,
$i^0=(x^+=\infty , x^-=-\infty )$ and $i^+=(x^+=\infty , x^-=0)$.}
\label{fig1}
\end{figure}

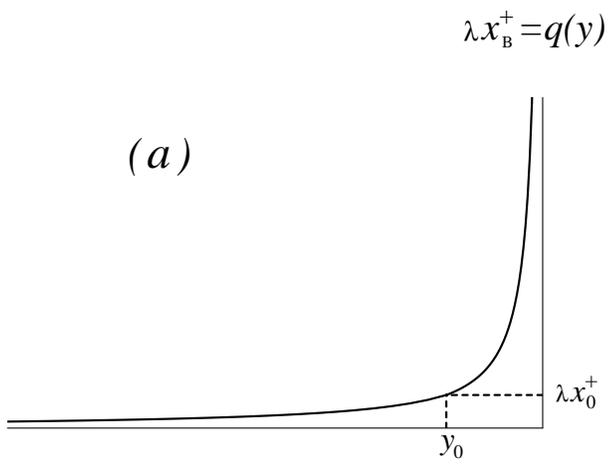
\begin{figure}
\caption{The boundary curve $\lambda x^+_B = q(y)$. In (a) the mass of the
infalling matter is $M=0.1 M_{cr}$ and in (b) it is $M=0.9 M_{cr}$. The
dimensionless coordinate $y = \lambda x^- + P_+/\lambda$ is plotted on the
horizontal axes.}
\label{fig2}
\end{figure}

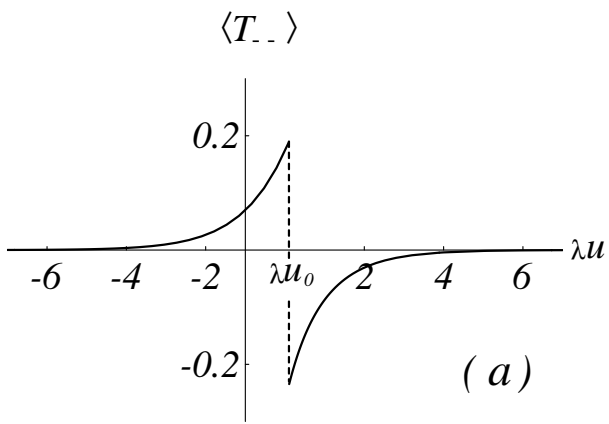
\begin{figure}
\caption{Quantum radiation on $\Im^+$. In (a) $M=0.1 M_{cr}$ and in (b)
$M=0.9 M_{cr}$. Here $\langle T_{--} \rangle $ is in units of
$\kappa \lambda^2 / 4$ (which is the value for Hawking radiation from 2D
black holes). In (a) $\lambda u_0 = 0.1$ and in (b) $\lambda u_0 = 2.3$.}
\label{fig3}
\end{figure}

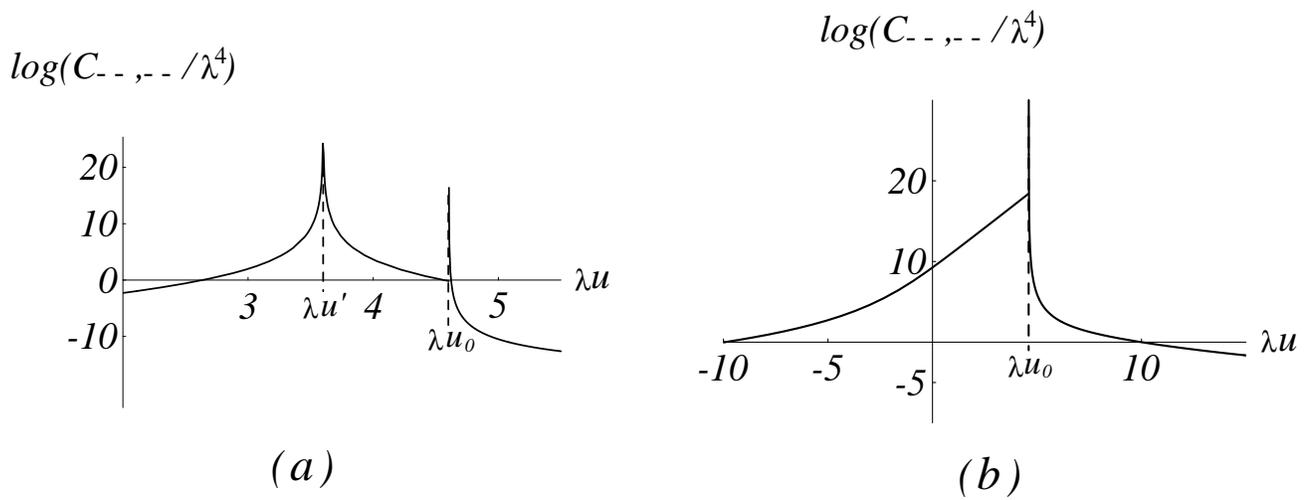
\begin{figure}
\caption{Log of the correlation function $C_{--,--}(u,u')$ for fixed $u'$.
In (a) we take $\lambda u' = 3.6$, and in (b) $\lambda u_0$
approaches $\lambda u_0 = 4.6$ from above.}
\label{fig4}
\end{figure}

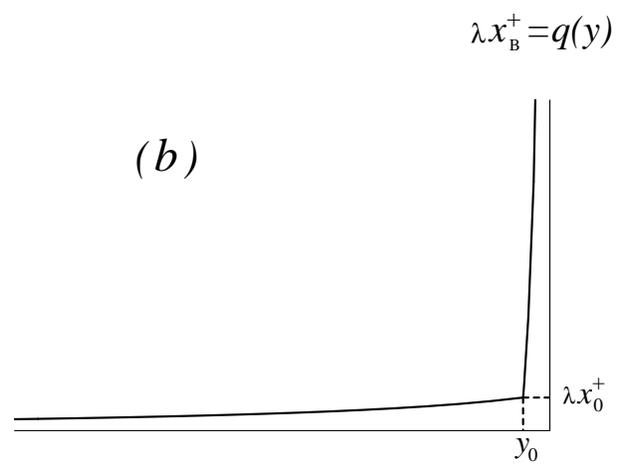
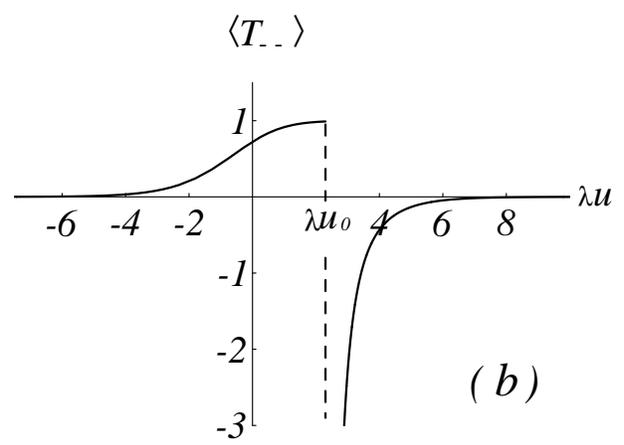
\end{document}